\documentclass{Interspeech}

\interspeechcameraready 

\title{Towards Bitrate-Efficient and  Noise-Robust Speech Coding \\ with Variable Bitrate RVQ}

\author[affiliation={1}]{Yunkee}{Chae}
\author[affiliation={1,2,3}]{Kyogu}{Lee}

\affiliation{Interdisciplinary Program in Artificial Intelligence}{Seoul National University}{}
\affiliation{Department of Intelligence and Information}{Seoul National University}{}
\affiliation{Artificial Intelligence Institute}{Seoul National University}{Republic of Korea}
\email{\{yunkimo95,kglee\}@snu.ac.kr}


\usepackage{comment}

\usepackage{mathtools}
\usepackage{amsfonts}
\usepackage{dsfont}

\usepackage{caption}
\usepackage{subcaption}

\usepackage{lipsum}

\usepackage{tikz}
\usepackage{nicematrix}

\usepackage{cite}
\usepackage{amsmath,amssymb,amsfonts}
\usepackage{algorithmic}
\usepackage{graphicx}
\usepackage{textcomp}
\usepackage{xcolor}
\usepackage{multirow}
\usepackage{booktabs}

\usepackage{hyperref}

\usepackage{etoolbox}

\AtBeginEnvironment{thebibliography}{%
  \setlength{\itemsep}{0pt}
  \setlength{\parskip}{0pt}
}

\let\oldthebibliography\thebibliography
\renewcommand{\thebibliography}[1]{%
  \oldthebibliography{#1}%
  \setlength{\itemsep}{-0.2pt}%
  \setlength{\parskip}{-0.2pt}%
}

\newcommand{\R}{\mathbb{R}}

\newcommand{\MC}{$M^\text{C}$}
\newcommand{\MFC}{$M^\text{C}_\text{FD}$}
\newcommand{\MV}{$M^\text{V}$}
\newcommand{\MFV}{$M^\text{V}_\text{FD}$}

\begin{document}

\maketitle

\begin{abstract}
Residual Vector Quantization (RVQ) has become a dominant approach in neural speech and audio coding, providing high-fidelity compression. 
However, speech coding presents additional challenges due to real-world noise, which degrades compression efficiency. 
Standard codecs allocate bits uniformly, wasting bitrate on noise components that do not contribute to intelligibility. 
This paper introduces a Variable Bitrate RVQ (VRVQ) framework for noise-robust speech coding, dynamically adjusting bitrate per frame to optimize rate-distortion trade-offs. 
Unlike constant bitrate (CBR) RVQ, our method prioritizes critical speech components while suppressing residual noise. 
Additionally, we integrate a feature denoiser to further improve noise robustness. 
Experimental results show that VRVQ improves rate-distortion trade-offs over conventional methods, achieving better compression efficiency and perceptual quality in noisy conditions.
Samples are available at our project page%
\footnote{%
  {\scriptsize\url{https://yoongi43.github.io/noise_robust_vrvq/}}%
}.

\end{abstract}

\section{Introduction}
Modern audio coding techniques, including speech codecs, aim to efficiently compress and transmit signals while preserving perceptual quality. 
Recent state-of-the-art neural audio codecs, such as SoundStream \cite{soundstream}, EnCodec \cite{encodec}, and Descript Audio Codec (DAC) \cite{dac}, have leveraged residual vector quantization (RVQ), enabling scalable bitrate control and high-quality compression. 
Moreover, their application to speech coding has further enhanced intelligibility and quality at lower bitrates, outperforming traditional codecs such as Opus \cite{valin2012opus} and EVS \cite{dietz2015evs}

However, speech coding differs from general audio coding in a fundamental way: Real-world speech signals often contain corruptions, such as background noise.
Most speech codecs are designed for clean input speech, making them highly vulnerable to quality degradation in noisy environments.
Therefore, integrating speech enhancement (SE) with speech coding is essential to ensure robust performance in real-world compression and communication systems.

Several studies have explored improving speech robustness in coding frameworks. 
\cite{lim2020robust, casebeer2021enhancing_into, jiang2022end_comm, omran2023disentangling, soundstream, huang2023twostage, li2025speech_icassp_new}
One approach applied a Siamese learning paradigm \cite{koch2015siamese} to train a noise-robust feature extractor, which was then used to condition a WaveNet \cite{van2016wavenet}, enhancing its ability to process degraded speech signals \cite{lim2020robust}. 
Another method introduced a compressor-enhancer encoder, utilizing a VQ-VAE-based autoencoder \cite{van2017neural_vqvae} in combination with a WaveRNN decoder \cite{kalchbrenner2018efficient_wavernn} to improve speech quality under noisy conditions \cite{casebeer2021enhancing_into}.
Jiang et al. \cite{jiang2022end_comm} jointly optimized speech enhancement and packet loss concealment, making it more robust in network transmission scenarios. 
Another approach investigated speech-background disentanglement in the compressed domain of neural audio codecs, further refining noise robustness at the encoding stage \cite{omran2023disentangling}.
SoundStream \cite{soundstream} also introduced controllable noise reduction using a FiLM (Feature-wise Linear Modulation) layer \cite{perez2018film}.
Additionally, a two-stage optimization procedure was proposed to achieve low distortion and high perceptual quality in noisy conditions, providing a theoretical foundation \cite{huang2023twostage}.
While these methods have significantly advanced speech enhancement and compression, most do not fully explore bitrate efficiency from a variable bitrate perspective, leaving room for further optimization in rate-distortion trade-offs.

Several of these studies \cite{soundstream, omran2023disentangling, huang2023twostage} have incorporated RVQ-based models for speech compression. 
Although RVQ has demonstrated strong performance, it operates under a constant bitrate (CBR) framework, allocating a fixed number of codebooks is each time frame of the latent vector, regardless of the complexity of the speech signal. 
This rigid allocation becomes particularly inefficient in noisy speech scenarios.
We hypothesize that noise-dominated frames (e.g., silent segments with background noise) require fewer codebooks, while speech-dominated frames (e.g., voiced segments) need more to preserve intelligibility and quality. 
However, constant bitrate RVQ (CRVQ) lacks the flexibility to adapt to these variations, resulting in wasted bitrate in noise-dominated frames and insufficient bit allocation for speech-heavy regions, ultimately reducing coding efficiency.

To address this limitation, we adopt variable bitrate RVQ (VRVQ), recently introduced in \cite{chae2024vrvq_icassp}. 
Unlike CRVQ, VRVQ dynamically adjusts the number of codebooks per frame using an \textit{importance map} trained to optimize the rate-distortion tradeoff. 
While importance map-based bitrate allocation has been explored in image compression \cite{li2018learning, mentzer2018conditional, li2020learning, lee2022selective}, its application to speech coding remains relatively unexplored.
This paper proposes a VRVQ-based approach for speech coding that mitigates excessive bit allocation to noise, achieving a superior rate-distortion trade-off compared to conventional CRVQ-based methods.
Additionally, we introduce a feature denoiser, inspired by image compression techniques \cite{cheng2022optimizing}, to further improve the noise robustness of the codec’s latent representation.

\begin{figure*}
    \centering
    \includegraphics[width=\linewidth]{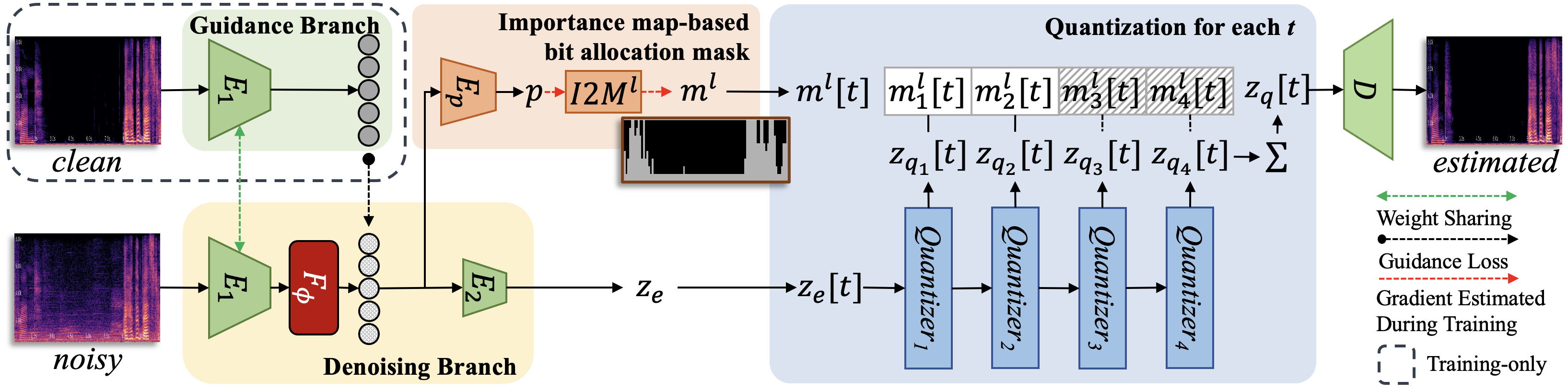}
    \caption{Overview of our framework.
    Our method leverages VRVQ \cite{chae2024vrvq_icassp} to jointly optimize the rate-distortion trade-off and denoising performance, drawing inspiration from \cite{cheng2022optimizing}. During training, both clean and noisy speech samples are fed into a shared encoder, while the feature denoiser is optimized using a feature guidance loss. 
    Note that the spectrograms in the figure are provided for illustration only; the actual input to the model is raw audio.}
    \label{fig:overall_framework}
\vspace{-5mm}
\end{figure*}

\section{Preliminaries}\label{sec:prelim}
This section provides a brief overview of RVQ, a fundamental component of state-of-the-art audio and speech codecs \cite{soundstream, encodec, dac, defossez2024moshi}, and VRVQ \cite{chae2024vrvq_icassp}, which introduces adaptive bit allocation for RVQ through an importance map.
\subsection{Residual Vector Quantization (RVQ)}\label{subsec:rvq}
RVQ is a hierarchical quantization method where an input latent representation is iteratively refined using up to $N_q$ quantizers.
Given input signal $x$, an encoder $E$ produces a latent representation $z_e\in\R^{D\times T}$, where $D$ is the latent dimension and $T$ is the number of downsampled time frames, and vector representation of each time $t$ is denoted as $z_e[t]$. 
For each time frame $t$, RVQ sequentially quantized the residuals as follows:
\vspace{-1mm}
\begin{equation}
    z_q[t]=\sum_{i=1}^{n_q}Q_i(r_i[t]),
\end{equation}
where $Q_i$ represents the $i$-th quantizer, and the residual is computed recursively as $r_i[t]=z_e[t]-\sum_{j=1}^{i-1}Q_j(r_j[t])$, starting with with $r_1[t]=z_e[t]$. Here, $n_q\leq N_q$ denotes the number of quantizers used.
The quantized output $z_q$ is then passed to the decoder for reconstruction.

\subsection{Variable Bitrate RVQ (VRVQ)}\label{subsec:vrvq}

VRVQ \cite{chae2024vrvq_icassp} extends RVQ by introducing adaptive per-frame bitrate allocation via an \textit{importance map}.
First, the encoder $E$ is decomposed into two parts: $E=E_2\circ E_1$, where $E_1$ extracts shared features, and $E_2$ performs encoding-specific processing. A learnable importance subnetwork $E_p$ takes the intermediate feature $E_1(x)$ and outputs an importance map $p\in(0,1)^T$, which determines the number of quantizers used per frame.

The bit allocation mask $m^l[t]\in\{0,1\}^{N_q}$ is derived from $p[t]$ using the importance map-to-mask (I2M) function:
\begin{equation}
    m^l[t]=\text{I2M}^l(p[t]):=[H^0(l\cdot p[t]),\dots,H^{N_q-1}(l\cdot p[t])]^\mathsf{T},
    \label{eq:i2m}
\end{equation}
where $H^k(s)$ is a Heaviside step function, defined as $H^k(s)=1$ if $s\geq k$, and $H^k(s)=0$ otherwise. 
Here, $l$ is a scaling factor for the bitrate control, preventing the importance map from producing static bit allocation patterns.
During training, $l$ is randomly sampled from a log-uniform distribution $U_{\log}(L_{\min}, L_{\max})$, while at inference time, it can be chosen based on the desired quality level.
The quantized output is then computed as:  
\vspace{-4mm}
\begin{equation}
    z_q[t] = \sum_{i=1}^{N_q} m_i^l[t] \cdot Q_i(r_i[t]).
    \label{eq:z_q}
\end{equation}  

Meanwhile, since the I2M function is non-differentiable, a smooth surrogate function of $H^k$ in Eq. \ref{eq:i2m} is introduced for backpropagation:
\begin{equation}
    f_\alpha^k(s) := \frac{1}{2\alpha} \log\left(\frac{\cosh(\alpha(s-k))}{\cosh(\alpha(-s+k+1))}\right) + \frac{1}{2},
    \label{eq:surrogate}
\end{equation}
where $\alpha\in\R_{>0}$ is a hyperparameter.
Then, the surrogate function of I2M, denoted as $\text{I2M}_\alpha$, is defined, and straight-through estimation is applied for backpropagation as follows:
\begin{equation}
    \text{I2M}_\alpha^l(p[t]):=[f_\alpha^0(l\cdot p[t]),\dots,f_\alpha^{N_q-1}(l\cdot p[t])],
    \label{eq:i2msoft}
\end{equation}
\vspace{-3mm}
\begin{equation}
    p[t]\mapsto \text{I2M}_\alpha^l(p[t]) + \text{sg}\left(\text{I2M}^l(p[t])-\text{I2M}_\alpha^l(p[t])\right),
\end{equation}
where $\text{sg}$ denotes the stop-gradient operation.

The rate loss $\mathcal{L}_R$ for training of importance subnetwork is defined as follows:
\vspace{-2mm}
\begin{equation}
    \mathcal{L}_R=\frac{1}{T}\sum_{t=1}^{T}E_p(\text{sg}(E_1(x)))[t].
    \label{eq:rate_loss}
\end{equation}
As in \cite{chae2024vrvq_icassp}, our models follow the methodology of DAC \cite{dac}, which employs adversarial training.
During the training, the rate-distortion (R-D) tradeoff loss $\mathcal{L}_{D}+\lambda_R\mathcal{L}_R$ is optimized.
Here, $\mathcal{L}_{D}$ is composed of several loss terms, following the formulation in \cite{dac}. 
It includes an L1 reconstruction loss computed on multi-scale log-mel spectrograms, a feature matching loss \cite{kumar2019melgan}, and adversarial losses derived from both a multi-period waveform discriminator (MPD) \cite{kong2020hifi} and a multi-resolution spectrogram discriminator (MSRD) \cite{jang2021univnet}. 
In addition, commitment loss, and codebook losses are incorporated, and $\lambda_R$ is a hyperparameter.

\begin{figure*}[!t]
    \centering
    \includegraphics[width=\linewidth]{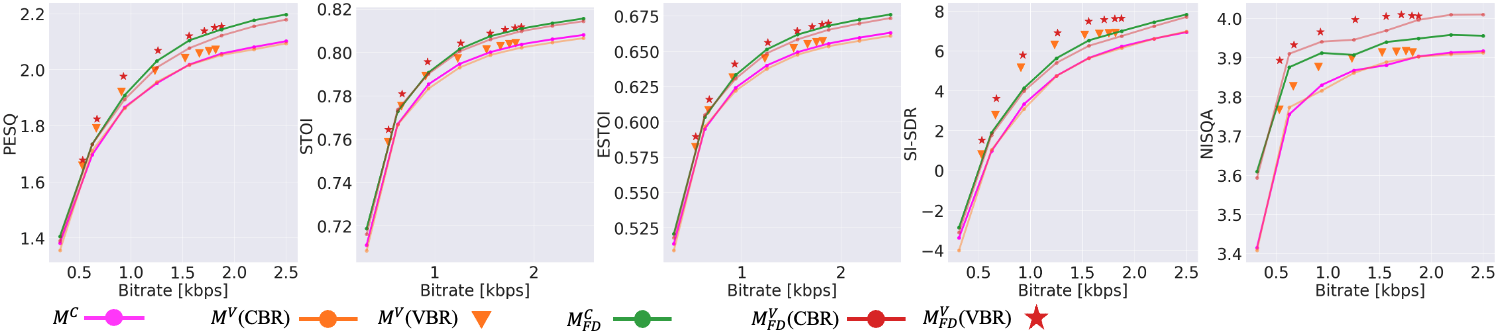}
    \caption{Results for the 16 kHz models. 
    In each subfigure, markers denote VBR inference outcomes at different scaling factors $l =$ 1.6, 2.6, 4.2, 6.9, 11.1, 18.1, 29.6, and 48, while the solid lines depict CBR inference results.}
    \label{fig:16k}      
    \vspace{-4mm}
\end{figure*}

\begin{figure}
    \vspace{-2mm}
    \centering
    \includegraphics[width=\linewidth]{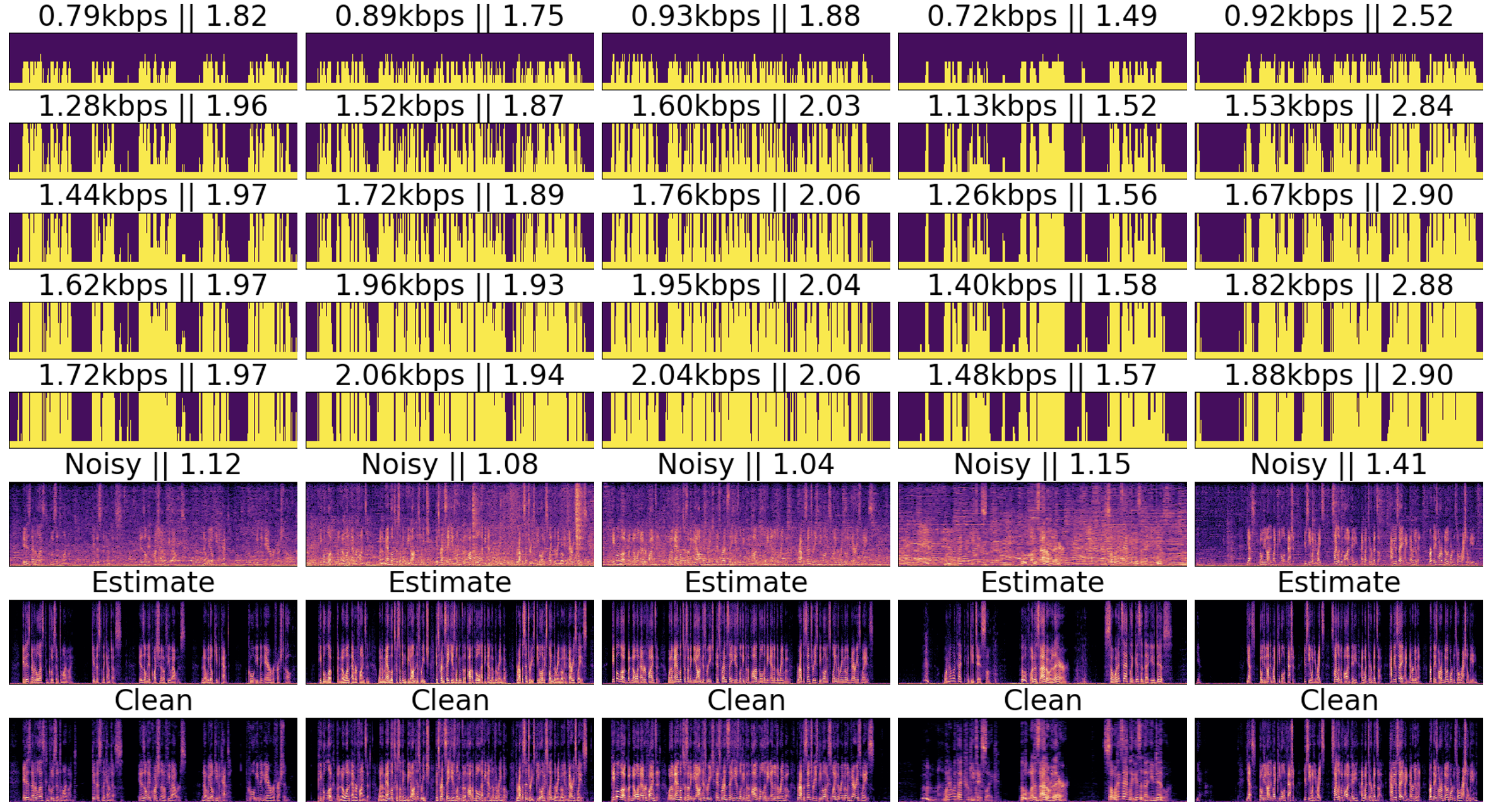}
    \caption{Illustration of the codebook allocation patterns for model \MFV. 
    Starting from the top row, the scaling factor $l$ are 4, 8, 12, 24, and 40, respectively. 
    Each panel is annotated with the corresponding average bitrate and its PESQ score.}
    \label{fig:impmaps}
    \vspace{-4mm}
\end{figure}

\section{Methodology}\label{sec:method}
In this section, we present our training framework leveraging VRVQ, as illustrated in Fig. \ref{fig:overall_framework}. 
Our approach consists of two key components: (1) \textit{Simple Mapping}, which directly trains a codec model to reconstruct clean speech from noisy input, and (2) \textit{Feature Denoising}, which further improves the joint optimization of speech enhancement and compression.
\subsection{Simple Mapping}\label{subsec:simple_mapping}
A straightforward approach to training the codec model $M$ is to directly map noisy speech $y$ to its clean counterpart $x$, using paired speech data $(x, y)$.
As a baseline, we train DAC \cite{dac} with constant bitrate RVQ (CRVQ), replicating the original DAC setup. 
We denote this model as \MC, where the codec is trained to estimate the denoised speech as $M^C(y)=\hat{x}$.
The training follows the same loss functions as in \cite{dac}, applied between $x$ and $\hat{x}$.
Additionally, we train the DAC with VRVQ \cite{chae2024vrvq_icassp}, denoted as $M^V$, following the same training scheme. 
As will be shown in Section \ref{sec:experiments}, the VRVQ model achieves significantly higher efficiency than the CRVQ baseline in terms of the rate-distortion trade-off, even under this simple framework.
\subsection{Feature Denoising}\label{subsec:feature_denoising}
To further improve joint speech denoising and compression, we introduce a feature-denoising process, inspired by the two-branch architecture used in image compression \cite{cheng2022optimizing}, as illustrated in Fig. \ref{fig:overall_framework}. 
Our approach comprises two stages:
\\
\textbf{1. Pre-training Stage.}
We begin by training the codec model exclusively on clean speech using a reconstruction-based approach for both CRVQ and VRVQ configurations.
In this stage, the models are optimized to ensure that $M^C(x)=\hat{x}$ and $M^V(x)=\hat{x}$.
\\
\textbf{2. Fine-tuning with Feature Denoiser.}
This phase builds upon the previous architecture by integrating a feature denoiser $F_\phi$.
All weights-except those associated with the $F_\phi$-are initialized from the pre-trained model obtained in the previous stage, and the complete model is fine-tuned on noisy speech data.
The feature denoising process is executed via two branches:
\begin{itemize}
    \item \textbf{Guidance Branch:}
    Clean speech $x$ is processed to extract intermediate features: 
    $z_\text{clean}=E_1(x)$.
    \item \textbf{Enhancing Branch:}
    Noisy speech $y$ is processed similarly to obtain: $z_\text{noisy}=E_1(y)$
\end{itemize}
The noisy features $z_\text{noisy}$ are passed through $F_\phi$ to generate a mask, refining the representation as follows:
\begin{equation}
    \hat{z}=\sigma(F_\phi(z_\text{noisy}))\odot z_\text{noisy},
\end{equation}
where $\sigma$ denotes a learnable sigmoid function inspired by the feature masking mechanism in Conv-TasNet \cite{luo2019convtasnet}.
We denote the codec models that incorporate a feature denoiser as \MFC and \MFV for CRVQ-based and VRVQ-based architectures, respectively.
To align the enhanced features with their clean counterparts, we introduce the feature guidance loss:
\begin{equation}
    \mathcal{L}_F=||z_\text{clean}-\hat{z}||_1.
\end{equation}
The final training objective is given by:
\begin{equation}
    \mathcal{L}=\mathcal{L}_D+\lambda_R\mathcal{L}_R+\lambda_F\mathcal{L}_F,
\end{equation}
where $\lambda_R$ and $\lambda_F$ are the weights for the rate and feature denoising loss terms, respectively.

\begin{figure*}
    \includegraphics[width=\linewidth]{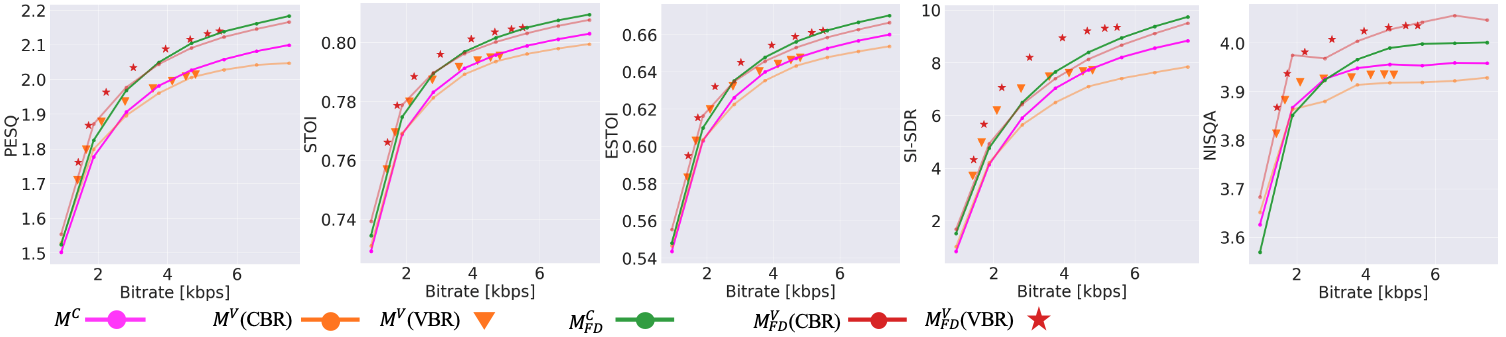}
    \caption{Results for 48k models.
    In each subfigure, markers denote VBR inference outcomes at different scaling factors $l =$1.6, 2.6, 4.2, 6.9, 11.1, 18.1, 29.6, and 48, while the solid lines depict CBR inference results.}
    \label{figs:48k}
    \vspace{-4mm}
\end{figure*}

\vspace{-2mm}
\section{Experiments}\label{sec:experiments}
\subsection{Model Architecture}
All RVQ-based codec models in our experiments are based on DAC \cite{dac} as described in Section \ref{sec:method}.
Each codebook uses 10-bit vector quantization, and the downsampling rate for time frame is set to 512. 
The maximum number of codebooks, denoted as $N_q$, is set to 8.
For the importance subnetwork, we adopt the same architecture as in \cite{chae2024vrvq_icassp}, which consists of five 1D convolutional layers with weight normalization \cite{salimans2016weightnorm} and Snake activation functions \cite{ziyin2020snake}. 
The final feature of the importance subnetwork has a channel dimension of 1, followed by a sigmoid activation function.
Our setup differs from \cite{chae2024vrvq_icassp} in that we prevent gradients from flowing from the importance subnetwork input back into the encoder, as described in Eq. \eqref{eq:rate_loss}.
We found that this stabilizes training, likely because it prevents the importance subnetwork from interfering with the original latent features in the encoder during training.

Our implementation of feature denoiser $F_\phi$ is based on the Mamba architecture \cite{gu2024mamba}, which leverages a state-space model with a selection mechanism.
Mamba has also been employed in recent state-of-the-art speech enhancement models \cite{chao2024semamba}. Accordingly, we implement $F_\phi$ using 10 bi-directional Mamba blocks.

\subsection{Experimental Setup}
\textbf{Dataset.} 
We employ the EARS-WHAM datasets \cite{richter2024ears, Wichern2019WHAM} in our experiments. 
The EARS dataset comprises anechoic chamber speech recordings spanning 22 emotion categories and 7 speaking styles. 
It provides paired noisy and clean speech samples, with 86.8 hours of audio for training, 1.7 hours for validation, and 3.7 hours for testing.
\\
\textbf{Training Configuration.} 
Models operating at 16 kHz are trained with a batch size of 32, while those at 48 kHz use a batch size of 16. 
For VRVQ models, 25\% of the samples in each mini-batch are processed using the full $N_q$ codebooks, whereas CRVQ models incorporate a quantizer dropout \cite{soundstream, encodec, dac} with a probability of 0.5 during training. 
The hyperparameter $\alpha$ in Eq. \eqref{eq:surrogate} is set to 2, and the loss weights are configured as $\lambda_R=3$ and $\lambda_F=0.1$.
During training, the scaling factor $l$ is sampled from a log-uniform distribution in the range $L_{\min}=0.8$ to $L_{\max}=48$.
Each model is trained for 200k iterations on a single NVIDIA RTX A6000 GPU, and the fine-tuning phase runs for 200k iterations.
\\
\textbf{Evaluation Metrics.}
We assess our models using multiple objective speech quality metrics. 
We evaluate intrusive metrics, including PESQ \cite{rix2001pesq}, STOI \cite{taal2011stoi}, ESTOI \cite{jensen2016estoi}, and SI-SDR \cite{le2019sisdr}, as well as the non-intrusive metric NISQA \cite{mittag2021nisqa}.
In addition, we report Bjøntegaard-Delta (BD)-rate \cite{bjontegaard2001calculation} for intrusive metrics using Akima interpolation \cite{herglotz2022beyond}—a widely adopted method in the video codec domain—to quantify the average differences between rate-distortion curves.

\newcommand{\tabhs}{@{\hspace{1.0mm}}}
\newcommand{\hmm}{2.0mm}

\begin{table}[t]
    \centering
    \renewcommand{\arraystretch}{0.6}
    \begin{tabular}{c@{\hspace{\hmm}}c@{\hspace{\hmm}}c@{\hspace{\hmm}}|
    c@{\hspace{\hmm}}c@{\hspace{\hmm}}c@{\hspace{\hmm}}c@{\hspace{\hmm}}} 
    \toprule
    \multirow{2}{*}{$f_s$}  & \multirow{2}{*}{Ref.} & \multirow{2}{*}{Tested} 
    & \multicolumn{4}{c}{BD rates $\downarrow$ (\%)} \\ \cline{4-7} \noalign{\vskip 1.5pt}
    &      &     & PESQ & STOI & ESTOI & SI-SDR  \\ \hline \noalign{\vskip 0.8pt}
    16 kHz & \MC & \MV &-9.44 &-3.87 &-4.58 &-18.39  \\ 
    16 kHz & \MC & \MFV &-18.22 &-14.77 &-14.25 &-26.33  \\ \noalign{\vskip 0.8pt}
    16 kHz  & \MFC & \MFV &-6.89 &-2.87 &-3.34 &-15.12  \\ 
    \noalign{\vskip 1pt} \hline \noalign{\vskip 0.8pt}
    48 kHz  & \MC & \MV &-8.76 &-7.23 &-6.68 &-16.11  \\
    48 kHz & \MC & \MFV &-27.31 &-25.01 &-21.90 &-32.57  \\ \noalign{\vskip 0.8pt}
    48 kHz  & \MFC & \MFV &-12.42 &-10.69 &-10.01 &-20.20 \\ 
    \bottomrule
    \end{tabular}
    \caption{BD-rate for intrusive metrics. Negative values indicate that the tested configuration achieves a lower bitrate for a given distortion level compared to the reference, reflecting improved rate-distortion performance.}
    \label{tab:bdrate}
    \vspace{-8mm}
\end{table} 

\subsection{Results}\label{subsec:results}
Fig.~\ref{fig:16k} illustrates the performance of models trained and evaluated on 16 kHz speech.
Since EARS-WHAM dataset consists of 48 kHz speech recordings, we downsampled the data to 16 kHz for training.
The resulting encoded latent vector has a sampling rate of 31 Hz.
In VBR inference mode, an extra transmission cost of $\lceil\log_2N_q\rceil=3$ bits per time frame is incurred (equivalent to an additional 0.093 kbps). 
Although these VRVQ models (\MV and \MFV) are trained in VBR  scheme using the importance map, they can also be executed in CBR mode by bypassing the importance map and using a fixed number of codebooks. 

As shown in Fig.~\ref{fig:16k}, training the simple mapping model using the VRVQ (i.e., \MV) achieves better performance than its CRVQ counterpart (\MC), confirming the effectiveness of VRVQ for simultaneous speech denoising and compression.
Moreover, incorporating feature denoiser (\MFV) yields further improvements over both \MV and its CRVQ counterpart with feature denoiser (\MFC).
It is also notable that across all metrics, inference in VBR mode consistently outperforms the CBR mode for the same VRVQ model. 

Fig.~\ref{figs:48k} presents the results for models trained and evaluated on 48 kHz speech data.
Here, the latent representation is downsampled to 93 Hz, which results in an additional transmission cost of 0.279 kbps. 
The overall trends remain similar to the 16 kHz experiments; however, the VRVQ model \MV exhibits slightly reduced performance at high bitrates compared to \MC.
In contrast, \MFV consistently achieves superior performance across all evaluated metrics compared to both \MC and \MFC.

Table~\ref{tab:bdrate} reports the BD-rate values for intrusive quality metrics. 
This table presents a numerical comparison between the reference and test models in each row, quantifying their relative performance.
For the VRVQ models, endpoints are defined by the inference results obtained using a fixed single codebooks and the full set of $N_q$ codebooks, respectivly, ensuring an accurate comparison with the reference CRVQ models.
Negative BD-rate values indicate that the tested configuration achieves a lower bitrate for a given distortion level relative to the reference, thereby reflecting improved rate-distortion performance.
Notably, all values in Table~\ref{tab:bdrate} are negative, indicating that the VRVQ models operate in a more bitrate-efficient manner across all metrics, with particularly significant gains in SI-SDR.
Moreover, \MFV clearly outperforms the reference CRVQ models for both 16 kHz and 48 kHz configurations.
Additionally, \MFV demonstrates greater efficiency than \MV, as evidenced by the lower BD-rate values when compared to \MV, with \MC as the reference.
It is particularly interesting that for 48 kHz models, even though \MV shows slightly lower performance at high bitrates compared to \MC, it still achieves a negative BD-rate when referenced to \MC, indicating overall bitrate efficiency improvements.

Fig. \ref{fig:impmaps} presents sample importance maps produced by the \MFV for noisy input audio. 
These visualization demonstrate that, despite the presence of high-frequency noise in the input audio, the importance subnetwork prioritizes bit allocation to regions corresponding to clean speech components. 
This targeted bit allocation effectively reduces redundant bitrate expenditure on noise,
thereby contributing to the overall performance gains observed.

\section{Conclusion} 
This paper presents a framework for simultaneous speech denoising and compression that leverages Variable Bitrate Residual Vector Quantization (VRVQ). 
Our experiments demonstrate that using the VRVQ scheme—compared to the conventional Constant Bitrate RVQ (CRVQ)—significantly enhances the performance of a simple mapping approach for converting noisy speech into clean speech.
Furthermore, the introduction of a feature denoiser yields additional performance gains, and its integration with the VRVQ scheme further enhances overall system performance.
In future work, we plan to develop more effective and lightweight training methods for feature denoising, with the ultimate goal of realizing a real-time system in VRVQ shceme.

\section{Acknowledgements}
This work was supported by the National Research Foundation of Korea (NRF) grant funded by the Korea government (MSIT) [No. RS-2024-00461617, 50\%], Information \& communications Technology Planning \& Evaluation (IITP) grant funded by the Korea government(MSIT) [No. RS-2022-II220320, 2022-0-00320, 40\%], [No.RS-2021-II211343, Artificial Intelligence Graduate School Program (Seoul National University), 5\%], and [No.RS-2021-II212068, Artificial Intelligence Innovation Hub, 5\%]

\begingroup
\tiny
\bibliographystyle{IEEEtran}
\bibliography{mybib}
\endgroup

\end{document}